      [1999/12/01 v1.4c Il Nuovo Cimento]
\documentclass{cimento}

\usepackage{graphicx}  
\title{Early GRB Optical and Infrared Afterglow Observations\ETC 
~with the 2-m Robotic Liverpool Telescope}
\author{A.~Gomboc\from{ins:x}\from{ins:y}\thanks{andreja.gomboc@fmf.uni-lj.si}, 
C.~G.~Mundell\from{ins:x}, C.~Guidorzi\from{ins:x}, 
A.~Monfardini\from{ins:x}\ETC, \\
C.~J.~Mottram\from{ins:x}, R.~Priddey\from{ins:z}, R.~J.~Smith\from{ins:x}, 
S.~Pak\from{ins:z}, I.~A.~Steele\from{ins:x}, N.~Tanvir\from{ins:z}, D.~Carter\from{ins:x}, 
S.~N.~Fraser\from{ins:x}, M.~F.~Bode\from{ins:x}, A.~M.~Newsam\from{ins:x} 
\atque M.~Hughes\from{ins:z}}

\instlist{\inst{ins:x} Astrophysics Research Institute, Liverpool John Moores University, UK
    \inst{ins:y} Faculty of Mathematics and Physics, University in Ljubljana, Slovenia
  \inst{ins:z} University of Hertfordshire, UK}
\PACSes{\PACSit{95.55.Cs }{Ground-based ultraviolet, optical and infrared telescopes}
\PACSit{98.70.Rz}{gamma-ray sources; gamma-ray bursts}}
\begin{document}

\maketitle

\begin{abstract} We present the first optical observations of a Gamma
Ray Burst (GRB) afterglow using the 2-m robotic Liverpool Telescope
(LT), which is owned and operated by Liverpool John Moores University
and situated on La Palma. We briefly discuss the capabilities of LT
and its suitability for rapid follow-up observations of early optical
and infrared GRB light curves.  In particular, the
combination of aperture, site, instrumentation and rapid response
(robotic over-ride mode aided by telescope's rapid slew and
fully-opening enclosure) makes the LT ideal for investigating the
nature of short bursts, optically-dark bursts,  and GRB blast-wave
physics in general. We briefly describe the LT's key position in the
RoboNet-1.0 network of robotic telescopes. We present the LT
observations of GRB041006 and use its gamma-ray
properties to predict the time of the break in optical light curve, a
prediction consistent with the observations.
\end{abstract}

\vspace*{-4mm}
\section{The Liverpool Telescope and RoboNet-1.0} Robotic operation of
the Liverpool Telescope (LT) enables an over-ride mode to be
automatically triggered when a GRB alert is received from GRB
Coordinates Network. Rapid response time ($\sim$1 min) combined with
2-m aperture and instrumentation (optical, near-infrared (NIR) camera and
spectrograph, for details see~\cite{ref:gom}) make the LT especially
suitable for the discovery and investigation of prompt optical/IR emission,
short-duration burst counterparts, and optically-dark long GRBs. Early-time
optical spectroscopy of bright bursts is also possible, and
statistical properties of GRBs and their afterglows will be a natural
consequence of the LT's fully-automated rapid response to GRB triggers.
  
Liverpool John Moores University (JMU) also leads the RoboNet-1.0 project,
the goal of which is to integrate three 2-m robotic telescopes - the
LT and two Faulkes Telescopes\footnote{Faulkes Telescopes, funded by the Dill Faulkes
Educational Trust, are almost
exact clones of the LT.  The Faulkes Telescope North (FTN) is situated
at Maui in Hawaii and has been operating since the end of
2003. Faulkes Telescope South (FTS), situated at Siding Spring,
Australia, achieved first light in September 2004.} - into a global
network optimised for round-the-clock rapid-response science, such as
exo-planet searches and GRB follow-up. 
RoboNet-1.0 is funded by the UK PPARC and includes members of
10 UK university teams in Cardiff, Exeter, Hertfordshire, Leicester,
Liverpool JMU, Manchester, Mullard Space Science Laboratory, 
Queen's University Belfast, St.~Andrews and Southampton. 
The capabilities of the telescope network, particularly in its
increased sky and time coverage, will greatly benefit GRB
detectability.

\section{Searching for Missing Optical Transients with the Liverpool Telescope}

To date, the question of `optically-dark' long GRBs has not been resolved~\cite{ref:jak},~\cite{ref:lam}.
Also, there have been {\em no} conclusive
detections of optical afterglows of short GRBs - the lack of
rapid and deep optical follow-up being a key problem.  We aim to
help 
answer whether these missing optical afterglows imply a
population of bursts that are (1) inherently dark; (2) dust absorbed;
(3) at high redshift or (4) just observationally overlooked.

\subsection{Long Bursts} Although few GRBs have optical detections in
the first ten minutes, two show rapid decay rates (3--5 magnitudes
over 10 minutes).  Given this rapid decline, it is feasible that the
50\% of so-called `optically-dark' long bursts may be easily detected
in sensitive, rapid observations with the LT. Alternatively, some
bursts may be optically dark due to obscuration or high redshift and
therefore identifiable only through infrared
photometry. Near-simultaneous optical and NIR imaging with the LT
during the first hour of the burst will identify optically-dark,
IR-bright low-redshift bursts that are highly obscured.  Similarly,
optically-dark, but time-delayed IR-bright transients will be
identified to z=10. Using the LT's NIR camera, SupIRCam, to obtain a
5-minute exposure will identify an OT with J $<$ 19 mag, equivalent to
detection of a bright afterglow at z=5, or H $<$ 18.5 mag. A 15-minute
exposure at H-band (H $<$ 19 mag) will identify a bright afterglow at
z=10, approximately 30 minutes after the burst.

\subsection{Short Bursts and New GRB Classes}
As no optical/IR counterpart of a short burst has yet been detected unambiguously, their origin - 
Galactic or extragalactic - and hence their energetics, progenitor and environment are all still 
unknown. The LT, with its combination of sensitivity and rapid response, is uniquely placed to detect these counterparts, which, if they exist, are predicted to be  3 -- 4 mag fainter than those of 
long bursts~\cite{ref:pan}; this  puts them beyond the reach of many current 
follow-up programmes.

\subsection{GRB Blast Wave Physics with the LT} We will use the LT to
study GRB blast-wave physics of optically-bright prompt/afterglow
emission to z $\sim$ 4 by obtaining early multicolour light curves and
spectroscopy of bright optical transients of long
bursts. These sources are expected to be easily observable and numerous,
thus enabling the early-time burst afterglows, energetics, evolution
and redshift distribution to be derived for a
statistically-significant sample. The use of GRBs as standard candles
will be tested.

\begin{figure}
\includegraphics[height=0.55\textwidth]{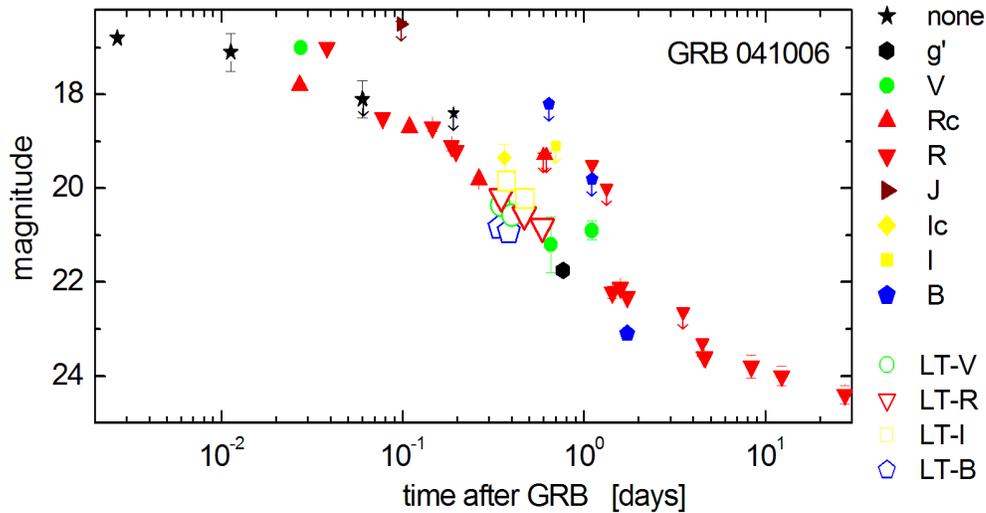}    
\caption{Optical light curves of GRB 041006, the first GRB afterglow observed with the LT. 
Data are taken from GCN Circulars: symbols mark detections and symbols+arrows mark upper limits. 
Data obtained with the LT are marked with open symbols.}
\label{fig1}
\end{figure}

\section{GRB 041006 with the Liverpool Telescope}

The first GRB afterglow observed with the LT was GRB
041006~\cite{ref:gal},~\cite{ref:dac}. Observations with the LT started 8.2 hours after the burst in
BVRI bands~\cite{ref:mon}, see Fig.\ref{fig1}, with calibration done
using LT standards and cross-checked against photometry reported in
Henden~\cite{ref:hen} and Fugazza et al~\cite{ref:fug}.  The temporal
power law decay with a slope of 1.2 is consistent with those of other
long-duration GRB afterglows.  

\subsection{Prediction of the Light Curve Break Time}

From the best-fit results of the average energy spectrum posted by
the HETE-2 team~\cite{ref:hete}
and assuming the validity of the Ghirlanda relationship
\cite{ref:ghi} between the peak energy of the $E\,F(E)$ spectrum
of the GRB in the rest frame, $E_{\rm p}^{\rm rest}$, and the collimation-corrected
total released energy in the rest-frame $1-10^4$~keV, $E_\gamma$,
we successfully predicted the time of the break in the
afterglow power-law decay.
The energy spectrum was fitted with a cut-off power law,
$N(E) = N E^{-\alpha} e^{-E/E_0}$, with the following best-fit
parameters: $E_0 = 100.2$~keV, $\alpha = 1.367$.
The peak energy of the $E\,F(E)$ spectrum turned out to be:
$E_{\rm p} = (2-\alpha) E_0 = 63.43$~keV, and the corresponding
value in the rest-frame therefore resulted: 
$E_{\rm p}^{\rm rest} = E_{\rm p} (1+z) = 108.6$~keV, where
we used the redshift $z=0.712$ provided by \cite{ref:fug}.

On one side, we know both the distance to the GRB and its energy
spectrum, so we measure directly its isotropic-equivalent
total released energy $E_{\gamma,{\rm iso}}$ (using the value for the 25--100~keV
fluence provided by HETE-2 team, i.e. $1.99\times10^{-5}$~erg cm$^{-2}$~\cite{ref:hete}):

\begin{equation}
 \label{e.gammaiso}
 E_{\gamma,{\rm iso}} = \frac{4\pi\,D^2_{\rm
      L}(z)}{(1+z)^2}\ \int_{1/(1+z)}^{10^4/(1+z)} E N(E) dE \quad
      \simeq\quad 8\times 10^{52}\ \textrm{erg}
\end{equation}

On the other side, by inverting the Ghirlanda relationship (eq.~2 in
\cite{ref:ghi}) from $E_{\rm p}^{\rm rest}$ we estimate the
collimation-corrected total released energy $E_\gamma$:
\begin{equation}
 \label{e.gamma}
 E_\gamma = 4.3 \ \Big(\frac{E_{\rm p}^{\rm rest}}{267~\textrm{keV}}\Big)^{1/0.706}
\ \times\ 10^{50}\ \textrm{erg} \quad\simeq\quad 1.2\times 10^{50}\ \textrm{erg}
\end{equation}
Combining the two, we estimate the collimation angle via the formula
$(1 - \cos{\theta}) = E_\gamma/E_{\gamma,{\rm iso}}$,
to be $\theta \simeq 0.055$~rad.
Assuming standard density $n\sim3$~cm$^{-3}$ and energy conversion efficiency
$\eta_\gamma=0.2$ (see \cite{ref:ghi}), the expected break time
(eq.~1 in \cite{ref:ghi}) results:
\begin{equation}
 \label{e.tjet}
t_{\rm jet,d} = (1+z)
  \Big(\frac{\theta}{0.161}\Big)^{8/3} \Big(\frac{E_{\gamma,{\rm
  iso},52}}{n \eta_\gamma}\Big)^{1/3} \approx 0.2\ \textrm{days}
\end{equation}

Despite
the uncertainty in the estimate of $E_p$, which was derived without WXM data,
the 
predicted value for the break time is 
consistent with the range 0.2--0.4~days inferred from the optical  afterglow light curve in 
Fig.~\ref{fig1}, see also \cite{ref:mal} and \cite{ref:mba}.

\section{Conclusions}

In view of many open issues in the field of GRBs and the current small
number of early-time optical observations within a few minutes or even
an hour after the GRB, it is clear that early multi-colour optical and
infrared photometry and spectroscopy can provide valuable information
on the nature of GRBs and their environments. With good GRB
localizations provided by Swift and other spacecraft, the Liverpool
Telescope will follow-up ~1 in 6 GRBs immediately following the alert
and provide continued monitoring of any fading afterglow. We will
obtain information on later time evolution of GRB light curves in
collaboration with other facilities, and all-sky coverage will be
enabled through RoboNet-1.0.

\acknowledgments
The Liverpool Telescope is funded via EU, PPARC and JMU grants and the benefaction of Mr
Aldham Robarts.
AG and CG acknowledge their Marie Curie Fellowships from the European Commission. 
CGM acknowledges the financial support from the Royal Society and MFB is grateful to the 
UK PPARC for provision of a Senior Fellowship.

\end{document}